\RequirePackage{fix-cm}
\documentclass[smallextended]{svjour3}       
\smartqed  
\usepackage{graphicx}
\usepackage{amsfonts}
\begin{document}


\title{The Bell experiment and the limitations of actors
}
\subtitle{Bell experiment}


\author{Inge S. Helland     
}


\institute{Inge S. Helland \at
              Department of Mathematics, University of Oslo \\P.O. Box 1053 Blindern, N-0316 Oslo, Norway\\
              Tel.: +47-93688918\\
              \email{ingeh@math.uio.no}           
           }

\date{Received: date / Accepted: date}

\maketitle

\begin{abstract}
The well known Bell experiment with two actors Alice and Bob is considered. First the simple deduction leading to the CHSH inequality under local realism is reviewed, and some arguments from the literature are recapitulated. Then I take up certain background themes before I enter a discussion of Alice's analysis of the situation. An important point is that her mind is limited by the fact that her Hilbert space in this context is two-dimensional. General statements about a mind's limitation during a decision process are derived from recent results on the reconstruction of quantum theory from conceptual variables. These results apply to any decision situation. Let all the data from the Bell experiment be handed over to a new actor Charlie, who performs a data analysis. But his mind is also limited: He has a four-dimensional Hilbert space in the context determined by the experiment. I show that this implies that neither Alice nor Charlie can have the argument leading to the CHSH inequality as a background for making decisions related to the experiment. Charlie may be any data analyst, and he may communicate with any person. It is argued that no rational person can be  convinced by the CHSH argument when making empirical decisions on the Bell situation. 
\keywords{Bell's theorem \and CHSH inequality \and conceptual variables \and limitation \and quantum foundation.}
\end{abstract}

\section{Introduction}
\label{intro}
Quantum mechanics is held by almost all physicists to be the most successfull theory ever deviced (although this assertion has been challenged \cite{RefJ1}). Nevertheless there has been and still are serious discussions about the interpretation of the theory. The relevant Wikipedia entry lists more than 16 different interpretations of quantum mechanics. Although some of these are related, this is obviously not a satisfying situation.

Much of the present discussion has centered around the Bell theorem \cite{RefJ2}: Quantum theory is inconsistent with local realism. In more concrete terms, the Bell inequalities, in particular the CHSH inequality \cite{RefJ3}, which is derived by a simple argument assuming local realism, can be violated by quantum mechanics. 

This has raised the question: Can the CHSH inequality also be violated by Nature, regardless of whether or not quantum mechanics is seen as an all-embracing true theory? Numerous experiments have been done to test this question, the first by Aspect et al. \cite{RefJ4}. These experiments have been criticized, and a list of possible loopholes have been identified \cite{RefJ0}. Finally. in 2015 several loophole-free experiments were performed \cite{RefJ5} \cite{RefJ6}, and the conclusion seems to be clear: \emph{There exist conditions under which the CHSH inequality is violated in practice.}

This has lead to new discussions: Should we abandon the hypothesis of locality, which seems to contradict relativity theory? (Using seemingly reasonable arguments, this is for instance claimed in \cite{RefJ7}.) Or should we in some sense or other question realism as a universal assumption?

In the present article, I will argue for a version of non-realism: From a general theorem on Hilbert space reconstruction it is proved that the mind of any actor will be limited in a given context. In concrete terms he is not able to have in his mind more than two relevant maximally accessible conceptual variables when making a decision, if these are essentially different and both related to his main thought. These terms are precisely defined. Applied to the Bell experiment, this is shown to imply that no actor is able to have all the assumptions behind the CHSH inequality in his mind in a context where he is to make decisions related to the experiment.

My arguments will rely on a general epistemic interpretation of quantum theory, advocated in the book  \cite{RefB1}. More details will be given below.

\section{The Bell experiment and the CHSH inequality}
\label{sec:2}

Two observers Alice and Bob are spacelikely separated at the moment when they observe. Midways between them is a source of entangled spin 1/2 particles, one particle in a pair is sent towards Alice, the other towards Bob. In concrete terms, the joint state of the two particles is given by
\begin{equation}
|\psi^0\rangle =\frac{1}{\sqrt{2}}(|1+\rangle |2-\rangle - |1-\rangle |2+\rangle ).
\label{Bell1}
\end{equation}
Here and in the following the spin component in any direction is normalized to $\pm 1$. In (\ref{Bell1}) $|1u\rangle$ means that the spin component of particle 1 in some fixed ($z$) direction is $u$, while $|2v\rangle$ means that the spin component of particle 2 in the $z$-direction is $v$. This state expresses that the total spin of the two particles is 0. One can imagine that these particles previously have been together in some bound state with spin 0.

Alice is given the choice between measuring the spin component of her particle in one of two directions $a$ or $a'$. If she measures in the $a$-direction, her response ($\pm 1$) is called $A$, and if she measures in the $a'$ direction, her response is called $A'$. Similarly, Bob can measure in one of two directions $b$ (giving a response $B$) or $b'$ (giving a response $B'$). The whole procedure is repeated $n$ times with different entangled particle pairs and with different directions/ settings chosen by Alice and Bob.

We now temporarily take the point of departure that all these response variables exist in some sense. This can be seen as an assumption on realism. At the very least, we will assume that this point of departure is meaningful for some observer or for a group of communicating observers.

Since all responses then are $\pm 1$, we then have the inequality
\begin{equation}
AB+A'B+AB'-A'B' = A(B+B')+A'(B-B')\le 2.
\label{Bell2}
\end{equation}
The argument for this is simply: $B$ and $B'$ are either equal to one another or unequal. In the first case, $B-B'=0$ and $B+B'=\pm 2$; in the last case $B-B'=\pm 2$ and $B+B'=0$. Therefore, $AB+A'B+AB'-A'B'$ is equal to either $A$ or $A'$, both of these being $\pm 1$, multiplied by $\pm 2$. All possibilities lead to $AB+A'B+AB'-A'B' =\pm 2$.

From this, a statistician will argue: Assume that we can consider $A, A', B$ and $B'$ as random variables, defined on the same probability space $(\Omega, \mathcal{F}, P)$. Then by taking expectations over the terms in (\ref{Bell2}), we find
\begin{equation}
E(AB)+E(A'B)+E(AB')-E(A'B')\le 2.
\label{Bell3}
\end{equation}

A physicist will have a related argument: Assume that there is a hidden variable $\lambda$ such that $A=A(\lambda), A'=A'(\lambda), B=B(\lambda)$ and $B'=B'(\lambda)$. The assumption that such a hidden variable exists, is called local realism in the physical literature. By integrating over the probability distribution $\rho$ of $\lambda$, this gives
\begin{equation}
E(AB)=\int A(\lambda) B(\lambda) \rho (\lambda)d\lambda
\label{Bell4}
\end{equation}
etc.. Thus by integrating term for term in (\ref{Bell2}), we again find (\ref{Bell3}).

 Of course the above two arguments are equivalent; it is just a question of using either the notation $(\omega , P)$ or $(\lambda ,\rho )$. There are different traditions here. These arguments are reviewed and discussed in detail by Richard Gill \cite{Ref18}.
 
  The inequality (\ref{Bell3}) is called the CHSH inequality after the authors of \cite{RefJ3}, and has been the source of much controversy. First, it is known that if we use quantum mechanics to model the above experiment, one can find settings such that the CHSH inequality is violated. Secondly, recent loophole-free experiments \cite{RefJ5} \cite{RefJ6} have shown that the CHSH inequality may be violated in practice.
  
  Thus the simple assumptions sketched above for (\ref{Bell3}) cannot hold.

\section{Briefly on the literature}
\label{sec:3}

There is a large physical literature around these questions. First, various authors have used rather advanced arguments claiming that Bell's theorem is wrong, and these arguments have each time been countered by Gill and collaborators (see for instance \cite{Ref12}). Much of the literature has recently been reviewed by Kupczynski \cite{Ref13}, and Marian Kupczynski has also arrived at his own conclusions there. I agree with him that a joint probability distribution of $(A, A', B, B')$ does not exist, hence that a joint probability distribution of the 4 variables $(AB,AB',A'B,A'B')$ does not exist. In a physical setting, Justo Pastor Lambare \cite{Ref17} has argued that this should imply that 4 different hidden variables $\lambda_i$ should be chosen in the equations corresponding to (\ref{Bell4}).

Another interesting argument has been put forward by Ilja Schmelzer \cite{RefJ7}. He argues that the key formula
\begin{equation}
E(AB|a,b)=\int A(a,b,\lambda ) B(a,b,\lambda)\rho(\lambda)d\lambda
\label{Bell5}
\end{equation}
follows from the logic of plausible reasoning (the objective Bayesian interpretation of probability theory) taken alone, and therefore that the violation of the CHSH inequality has as a consequence that Einstein causality (in essence the assumption of nonlocality) has been violated.

To counter this last argument, we must go somewhat into the logic of plausible reasoning. Applied statistics is based upon a large number of propositions about parameters, and we may assume that the set of all these propositions form a Boolean algebra. The tradition in statistics is to associate such Boolean algebras with set theory, and in fact this association can be made precise. Mathematically, a Stone space is a compact totally disconnected Hausdorff space; these details are not too important. But Stone's representation theorem \cite{Ref14} says that every Boolean algebra $B$ is connected to a Stone space $S(B)$ in the following sense: The topology on $S(B)$ is generated by a (closed) basis consisting of all sets of the form $\{x\in S(B) | b\in x\}$. Then every Boolean algebra $B$ is isomorphic to the algebra of subsets of its Stone space $S(B)$ that are both closed and open.

The problem is to which extent one can associate probabilities to all such propositions about parameters. Here the answer depends on which school in statistical inference you belong to. An extreme Bayesian will be willing to assign probabilities to all propostions. Most statisticians find themselves in an in-between position, for some statements they can associate probabilities, epistemic probabilities, for some statements they can not. Sometimes, like when the problem in question has some symmetry and we can associate a prior to the right invariant measure of a transitive group $G$, one can act as a Bayesian, but often this attitude is not possible. In this article I will also support the frequentist tradition in statistics, where prior probabilities of propositions are not necessarily assumed to be available. Taking such an attitude, statements like (\ref{Bell5}) are not automatically true. In fact, we will argue below that all expectations and every statistical analysis should be taken from the point of view of an observer/ actor or from the point of view of a group of communicating actors. This point of departure is central for all the different arguments in the book \cite{RefB1}, and it will be crucial for the discussion that I will make below.

\section{An epistemic approach towards quantum theory}
\label{sec:4}

In the literature, many different interpretations of quantum theory are given. Some of these emphasize the epistemological aspect of the quantum state, which I will also do here. Among these, one could mention QBism \cite{Ref2}, founded by Christopher Fuchs, and Carlo Rovelli's Relational Quantum Mechanics \cite{Ref3}. The latter relies upon 2 hypotheses: 1) All systems for describing systems relative to an observer are equivalent. 2) Quantum mechanics provides a complete scheme for description of the physical world. A general physical theory is a theory about the state that physical systems have, relative to each other. In particular, an observer may be such a system.

In \cite{RefB1} a general epistemic view upon quantum theory is advocated. The basis is a set of conceptual variables connected to some agent/observer or shared by a group of communicating observers. Some of these variables may be given numerical values through some experiment; these are called accessible variables or epistemic conceptual variables (e-variables). An example may be the spin component $\theta^a$ of a particle in some given direction $a$. (In the Bell experiment setting above, I have used the notation $A=\theta^a$ etc.) Other variables are inaccessible, can not be given values. An example may be the full spin vector $\phi$ of a particle. Variables such as $\theta^a$ which can not be extended without loosing their accessability property, are called maximally accessible. Based upon this view, ontological aspects of the quantum state may also be considered \cite{Ref71}.

One must distinguish sharply between conceptual variables attached to a single actor and conceptual variables attached to a group of actors. According to Zwirn \cite{RefZ}, see below, only the first variables have a primary role to play in our description of the world. But when making decisions, and when arriving at joint descriptions after having communicated,  both kinds of variables will be important.

In the views developed in \cite{RefB1}, quantum theory may be based upon concentrating on such conceptual variables. In agreement with N. David Mermin \cite{Ref25} the only `real things' in physics are events. In the discrete case an event is given in some context by `$\theta=u$', where $\theta$ is a conceptual variable, and $u$ is one of its values. It is argued in \cite{RefB1} that these variables either should be connected to the mind of a single actor or the joint minds of a group of communicating actors. And they should always be associated with some context and with a concrete physical situation.

It is important that my conceptual variables are connected to a concrete context. By the well-known Kochen-Specker theorem it is impossible to assign simultaneously, noncontextual definite values to all (of a finite set of) quantum mechanical observables in a consistent manner.

The following result is developed in \cite{RefB1} and improved in \cite{Ref50}: For every maximally accessible discrete e-variable $\theta$, varying on some space $\Omega_\theta$, on which a transitive group $G$ can be defined, there corresponds under weak condition a unique operator $A^\theta$ defined on some common Hilbert space $\mathcal{H}$, and to every question `What is the value of $\theta$ if measured?' together with a sharp answer `$\theta=u$' there is a vector, an eigenvector of $A^\theta$ with eigenvalue $u$. 

More precisely, in \cite{Ref50} (cp. also \cite{RefB1}) the following is proved:
\bigskip

\textbf{Theorem 1.}
\textit{Consider a situation where there are two maximally accessible conceptual variables $\theta$ and $\eta$ in the mind of an actor or in the minds of a communicating group of actors. Make the following assumptions:}

\textit{(i) On one of these variables, $\theta$, there can be defined group actions from a transitive group $G$ with a trivial isotropy group and with a left-invariant measure $\rho$ on the space $\Omega_\theta$.}

\textit{(ii) There exists a unitary irreducible representation $U(\cdot)$ of the group $G$ defined on $\theta$ such that the coherent states $U(g)|\theta_0\rangle$ are in one-to-one correspondence with the values of $g\in G$, and hence with the values of $\theta$.}

\textit{(iii) The two maximally accessible variables $\theta$ and $\eta$ can both be seen as functions of an underlying inaccessible variable $\phi\in\Omega_\phi$. There is a transformation $k$ acting on $\Omega_\phi$ such that $\eta(\phi)=\theta(k\phi)$.}

\textit{Then there exists a Hilbert space $\mathcal{H}$ connected to the situation, and to every accessible conceptual variable there can be associated a unique symmetric operator on $\mathcal{H}$.}
\bigskip

A simple assumption implying the technical condition (ii) is given in \cite{Ref50}, and this technical condition can be shown to hold  in the spin 1/2 case.

Condition (iii) is of particular interest. Two conceptual variables satisfying this condition are said to be \emph{related}. When it is impossible to find an underlying variable $\phi$ such that this condition holds, we say that $\theta$ and $\eta$ are \emph{essentially different}.

In particular, these conditions hold for pairs of components $\theta^a$ and $\theta^{a'}$ of a spin 1/2 particle, where $\phi$ is the inaccessible spin vector, and $k$ is a particular rotation of $\phi$. (We can take $k$ as a $180^o$ rotation around the midline between the directions $a$ and $a'$.) Then the corresponding Hilbert space $\mathcal{H}$ is two-dimensional, and every question `What is the value of $\theta^a$?' together with an answer `$\theta^a =u$', where $u=\pm 1$, corresponds to a unique unit vector in $\mathcal{H}$. In fact, here, every unit vector in $\mathcal{H}$ has an interpretation in the form of a question-and-answer pair. This has recently been generalized to the case of several such questions by H\"{o}hn \cite{Ref30} and H\"{o}hn and Wever \cite{Ref20}.

The assumption that there can be defined a transitive group acting upon $\theta$ is crucial. It can easily be satisfied when the range of $\theta$ is finite or is the whole line ${\mathbb{R}}^1$, but may be more difficult when $\theta$ is a vector.

Thus Theorem 1 can be used in a new foundation of parts of quantum theory, but this foundation may then in practice first be limited to finite-valued or scalar variables, and where these are maximally accessible. After this, operators for other accessible conceptual variables can be found by looking at them as functions of a maximal variable, and then making use of the spectral theorem for the operator corresponding to this variable. Note also that in a situation where we have several independent (scalar) accessible conceptual variables, an operator corresponding to the vector composed of all these may be defined by taking tensor products. 

Experiments to measure an accessible conceptual variable $\theta$ are mostly seen as perfect in the quantummechanical literature, measurement apparata are seen as completely accurate. In practice, the apparata give inaccurate data $x$ which are connected to $\theta$, and this inaccuracy may be modeled by a statistical model $p(x|\theta)$. The observer/ experimentalist may then give an estimate $\widehat{\theta}$ from the data $x$. From the point of view of Convivial Solipsism \cite{RefJ8}, this estimate may be seen as the result of measurement perceived by the consciousness of the observer, taken with respect to later decisions as the true resulting value of $\theta$ if the current experiment is accurate enough.

When the result of the experiment is discrete, as in the Bell experiment case, it is often not problematic to regard the estimate from the data as the `true' value in some sense (but, and this is important, again connected to an actor or to a group of communicating actors). This will be assumed in the following. In one run, $\theta^a =A$ may be seen as a fixed value, but seen from the point of view of many repeated runs, they are random variables. But they are random variables in an epistemic sense, their probablity distributions are epistemic probabilities. (For a discussion of this concept in a statistical setting, see \cite{RefB1}). And, I repeat: The epistemic conceptual variables are connected to an actor, for instance Alice in the Bell experiment, or to a group of communicating actors. Later I will let Alice communicate with a more knowledgeable actor Charlie.

\section{The conditionality principle}
\label{sec:5}

The theory of statistical inference relies on certain principles, one of these is the conditionality principle. The principle was first proposed by David Cox \cite{Ref15} in 1958, based on a very simple example: Suppose that I, as a chemist has made measurements on some material, and I want my measurements to be analysed by a laboratory before I make a simple statistical treatment of the results. I have the choice between two laboratories, one in New York and one in San Fransisco. I decide to toss an unbiased coin to make a decision between the labs.

In principle one might imagine that my final statistical analysis should be based on the whole epistemic process, including the coin tossing. But David Cox made the very reasonable assumption that one should condition the statistical analysis on the result of this coin toss.

He then made a very bold generalization of this example: Let a data variable $z$ have a distribution that is independent of the parameters of the experiment that we want to analyze. Then $z$ is called ancillary.  The general conditionality principle then says: \emph{One should condition the statistical analysis on the value of any ancillary data variable.}

One can discuss the principle in this generality; in fact, I have done so \cite{Ref16} several years ago. But my point is now: The conditionality principle should be used on the settings chosen by Alice and Bob in the Bell experiment in their own statistical analysis of their data. The $z$ is then, for each of them, the settings chosen (which in some discussions are thought about as the results of some coin tosses.)

\section{Convivial solipsism}
\label {sec:x}

A new philosophy, convivial solipsism, which may help to understand the basis of quantum mechanics, a philosophy which also may be linked to my own ideas, was recently proposed by Zwirn \cite{RefZ}.

In general, solipsism is a philosophy with many variants. It is based on the view that everything that we can know for sure by our mind is connected to this mind. My mind is an autonomous separate world. The convivial variant also recognizes that other people have their minds, and thus have sure statements connected to their minds. And communication between different people is possible. People that have communicated and agreed on certain questions may be seen as a new unit, a new world, with respect to these questions. In a macroscopic context this may be linked to a theory of making decisions.

\section{Born's formula}
\label{sec:6}

In \cite{RefB1} the Born rule is formulated as follows: Assume in general two maximally accessible conceptual variables $\theta^a$ and $\theta^b$. Let, for some observer or group of observers the event $\theta^a =u$ correspond to the ket vector $\psi^a$ and the event $\theta^b = v$ correspond to the ket vector $\psi^b$ in some common Hilbert space $\mathcal{H}$. Then
\[P(\theta^b =v |\theta^a =u)=|\langle\psi^b |\psi^a \rangle |^2.\]

In the Bell experiment, if $A=\theta^a$ is Alice's response and $B=\eta^b$ is Bob's response, and assuming an actor for which both $A$ and $B$ are meaningful, can be related to the same Hilbert space, we find from this
\[P(B=\pm 1|A=+1)=(1\pm\mathrm{cos}(a,b))/2,\]
and assuming in addition that $P(A=-1)=P(A=+1)=1/2$, this gives $E(AB)=-\mathrm{cos}(a,b)$. From this again follows that according to quantum mechanics, the CHSH inequality may be violated, for instance $a\sim 0^o$, $a'\sim 90^o$, $b\sim 225^o$ and $b'\sim 135^o$ gives
\begin{equation} 
E(AB)+E(AB')+E(A'B)-E(A'B')=2\sqrt{2}.
\label{notBell}
\end{equation}

In \cite{RefB1} the Born rule was derived by making three assumptions which qualitatively may be formulated as: 1) The variable $\theta^a$ (here $A$) is maximally accessible; 2) The so-called likelihood principle from statistics holds; 3) The observer is either himself perfectly rational, or has ideals which can be thought of in terms of an artificial, perfectly rational actor, where rationality is given by the Dutch book principle. Of course an additional assumption behind (\ref{notBell}) is that each of the conceptual variables $AB$, $AB'$, $A'B$ and $A'B'$, taken separately, make sense to the actor(s) in question.

\section{A mathematical theory of human decisions}
\label{sec:7}

Let the person A be in a decision process. He has the choice between the prospects $\pi_1,...,\pi_r$. Introduce a decision variable $\xi$ taking $r$ values: $\xi = k$ if and only if $\pi_k$ is to be realized ($k=1, ...,r$). This is a variable in the mind of A.

In general, A is in some context at the time t when he is to take his decision. In this context and at this time he can have several variables in his mind: $\theta, \xi, \lambda, ...$. I will call these conceptual variables. If a variable $\lambda$ can be given some value at a future time, I will say that $\lambda$ is accessible.

Say that a conceptual variable $\theta$ is `less than or equal to' the conceptual variable $\lambda$ if $\theta = f(\lambda)$ for some function $f$. This defines a partial ordering both among all conceptual variables and also among the accessible conceptual variables.

I will assume that if $\lambda$ is accessible, and $\theta =f(\lambda)$, then $\theta$ is also accessible.

Now to a main assumption of my model: I will assume that all of the conceptual variables in the mind of A, or some of them, can be seen as functions of an underlying inaccessible $\phi$, belonging to the subconsciousness of A. As such, $\phi$ can never be known by A, nor by any other person. Some intelligent person, knowing A, having observed him over some period, and knowing some practical psychology, may perhaps find a rough estimate of $\phi$.

In the case where the accessible conceptual variables are spin components of a particle, we can let $\phi$ be the inaccessible spin vector.

In general from this and from Zorn's lemma, maximally accessible conceptual variables (according to this partial ordering) always exist, since $\phi$ can be seen as an upper bound for a set of accessible conceptual variables. Trivially, the spin components of a particle are maximally accessible.

\section{The data analysis made by Alice}
\label{sec:8}

Assume that a Bell experiment has been done. Before she has any contact with Bob, Alice has a list of $n$ data from herself, settings $a$ or $a'$ and corresponding responses $A$ or $A'$. 

 Now by the conditionality principle, her analysis should be conditional, given her setting, either $a$ or $a'$. Assume that she first concentrates on the runs with setting $a$ and a corresponding response $A$. For example one may concentrate on the runs where $A=+1$, an event which corresponds to a unique ket vector $\psi^a$ in her two-dimensional Hilbert space $\mathcal{H}$. In fact the argument below will hold for any state of knowledge about the responce $A$. We will be particularly interested in the mixed state determined by $P(A=-1)=P(A=+1)=1/2$.
 
 By Theorem 1, the Hilbert space describing Alice's mind in the situation can be constructed from two maximally accessible conceptual variables.
 
With this as a background, one can imagine two scenarios. First, she may also have in mind the possible response $A'$ from her other setting $a'$. Then from a general theory of a mind's limitation, discussed in the next section, she is not able to think of any more conceptual variables. In particular, the possible responses made by Bob can not be addressed by her in this state, and to her then, the argument leading to (\ref{Bell2}) and (\ref{Bell3}) does not make any sense. Thus she has no opinion about the validity or not of the CHSH inequality.
 
 The other scenario is that she does not think of $A'$ at all, but concentrates her mind on the possible responses made by Bob. Let us assume that she knows some quantum mechanics, in particular the Born rule. Then she may use this rule to calculate $E(B|A=+1)$ and $E(B'|A=+1)$ from the known settings $b$ and $b'$, and assuming $P(A=-1)=P(A=+1)=1/2$
 she can calculate $E(AB)$ and $E(AB')$, that is, the first and the third term in (\ref{Bell3}). But there is no way in which she can get any information on the second and fourth term. Thus Alice is not able to give any meaning to the left hand side of the CHSH inequality, and this inequality might well be violated if we only are allowed to take into account the information posessed by Alice. Again she can have no opinion on the CHSH inequality.

This conclusion is of course the same if the setting chosen by Alice is $a'$. And a completely similar discussion can be made seen from Bob's point of view. The conclusion is that the simple reasoning leading to the CHSH inequality can not be made meaningful to either of these observers at a stage where they only know their own responses.

For practical experiments, not assuming quantum theory, one might perhaps imagine some other probability model doing the job that Born's formula did above, but the problem with missing information about two terms on the lefthand side of the CHSH inequality is the same.

Note that in spite of all this, Alice may well be very intelligent. Her Hilbert space relevant also to some specific other context may be of the form $\mathcal{H}\otimes\mathcal{K}$, where $\mathcal{H}$ is her two-dimensional Hilbert space connected to the Bell experiment, and $\mathcal{K}$ is a fairly big Hilbert space connected to the other context.

\section{A general theory of a mind's limitation}
\label{sec:9}

Take as a point of departure any experimental situation or decision situation and an observer/ actor $O$ in this situation. $O$ will have in his mind several conceptual variables connected to the situation. Assume that two of these are maximally accessible. I will assume that the conditions of Theorem 1 hold. 

According to Theorem 1 the situation can then be described by a Hilbert space $\mathcal{H}$. In \cite{RefB1} the corresponding theorem was proved with an extra condition connected to an epistemic process, but it was shown in \cite{Ref50} that this extra condition is in fact unnecessary. This is an important observation. It implies that the conclusion of the theorem applies to any decision situation, and that the conceptual variables involved may be decision variables or underlying variables that somehow influence decisions.

Another qualification is the condition (iii) in Theorem 1. If $\theta$ and $\eta$ together satisfy this condition, I will say that they are \emph{related}. It $\theta$ and $\eta$ can not be related in this way, we say that they are \emph{essentially different}.

In this Section I will assume that the technical conditions behind Theorem 1 are satisfied. For simplicity assume first also that the Hilbert space has a finite dimension $d$. Look at one of $O$'s maximally accessible variables $\theta$, say taking the values $u_1,..., u_d$. This has a unique operator $A^\theta$ connected to it. First, it follows from Theorem 4.5 i \cite{RefB1} that the eigenspaces of $A^\theta$ are one-dimensional, and it follows from Theorem 4.4 in \cite{RefB1} that the eigenvalues of $A^\theta$ are just $u_1,...,u_d$. And each event $\theta=u_i$ corresponds to a unique unit vector $|\psi^i\rangle$ in $\mathcal{H}$, the eigenvector of $A^\theta$ giving the eigenvalue $u_i$.

It is relevant to look at a theory of decisions as described by Yukolov and Sornette \cite{Ref35}. This reference is only one of a series of papers written on Quantum Decision Theory by the same authors. Similar conclusions may be made by taking as points of departure cognitive models, as developed in \cite{Ref21} and \cite{Ref23}.

In agreement with \cite{Ref35} and \cite{RefB1} let us assume for simplicity that the current state of mind of $O$ is given by a ket vector $|\psi\rangle$ in $\mathcal{H}$. He is going to make a decision, and his possible prospects $\pi_j$ are each represented by ket vectors $|\pi_j\rangle$ in $\mathcal{H}$. Then, according to the Born rule, his probability of making decision $\pi_j$ is given by
\begin{equation}
 p(\pi_j) = |\langle\psi |\pi_j \rangle |^2 .
 \label{prospect}
 \end{equation}

More generally, if his current state is given by a density matrix $\rho$, the probability of making decision $\pi_j$ is given by $p(\pi_j)=\langle \pi_j |\rho \pi_j \rangle$. 

The concept of permissibility is defined in the Appendix.
\bigskip

\textbf{Theorem 2}
\textit{Assume that the individual $O$ has two related maximally accesible variables $\theta$ and $\eta$ in his mind. Then $\eta(\phi)=\theta(k\phi)$ for an inaccessible variable $\phi$ and a transformation $k$ of $\Omega_\phi$. Assume that a group $K$ of transformations of $\Omega_\phi$ can be found such that $k\in K$ and $\theta(\cdot )$ is permissible with respect to $K$.}

\textit{In this situation $O$ can not simultaneously have in mind any other maximally accessible variable which is related to $\theta$, but essentially different from $\eta$.}
\bigskip

\textit{Proof.} According to Theorem 1, two different related maximally accessible variables, say $\theta$ and $\eta$ in the mind of $O$ will determine his Hilbert space $\mathcal{H}$ in the given context. And from this, all other conceptual variables in his mind will be associated with selfadjoint operators. Assume that one of these, say $\eta'$, essentially different from $\eta$, is maximally accessible. Then a different alternative theory could have been developed from $\theta$ and $\eta'$, giving a Hilbert space $\mathcal{H}'$. We will show that this leads to a contradiction.

Since $\theta$ and $\eta'$ are related, there is a transformation $k'$ such that $\eta'(\phi)=\theta(k'\phi)$. Extend if necessary $K$ to a group $K'$ such that $k'\in K'$. We need the following:
\bigskip

\textbf{Lemma 1} \textit{In this situation $\theta(\cdot )$ is permissible with respect to the group $K'$.}
\bigskip

\textit{Proof of Lemma 1.} We have that $\theta(\phi_1 )=\theta(\phi_2)$ implies $\theta(k'\phi_1 )= \theta(k'\phi_2)$, and further $\theta(h\phi_1 )=\theta(h\phi_2)$ for all $h\in K$. It follows from this that $\theta(hk'\phi_1 )=\theta(hk'\phi_2)$ and $\theta(k'h\phi_1 )=\theta(k'h\phi_2)$ for all $h\in K$. A similar property holds for all group elements that can be written as a finite product of $k'$ and elements in $K$. But these products generate $K'$.
\bigskip

\textit{Proof of Theorem 2, continued.} From Theorem A1 in the Appendix it follows that $A^\theta =V(k)A^\eta V(k)^\dagger$ and $A^\theta =V(k')A^{\eta'} V(k')^\dagger$. But this implies that $A^\eta =V(k^{-1}k')A^{\eta'} V(k^{-1}k')^\dagger$, and using Theorem A1 again, we conclude that $\eta$ and $\eta'$ are related, which leads to a contradiction with the assumptions made.

\qed

\bigskip

Theorem 2 also holds for continuous variables if one can assume that the technical condition (ii) of Theorem 1 holds.
\bigskip

Go back to the previous section, the situation of the actor Alice. Her Hilbert space is two-dimensional, and her current state $|\psi\rangle$ can be imagined to be determined by the event of the type $A=+1$. Her Hilbert space may be reconstructed by considering in addition one other  binary variable $\eta$. As discussed there, this variable may be either  her other response $A'$, or one of Bob's responses $B$ or $B'$. There are also other possibilities. But she is not able to, when making decisions, have other, essentially different variables in her mind. To prove this from Theorem 2, we concentrate on the case $A=\theta^a$ and $B=\theta^b$, where we need the following:
\bigskip

\textbf{Lemma 2} \textit{Let $K$ be the group of rotations of the spin vector in the plane spanned by $a$ and $b$. Then both the components $\theta^a$ and $\theta^b$ are permissible with respect to $K$, and $\theta^a$ and $\theta^b$ are related.}
\bigskip

\textit{Proof.} Permissibility follows essentially as in Proposition 2 in \cite{Ref50}. $\theta^a$ can obviously be transfered into $\theta^b$ by a suitable rotation in this plane. 

\qed

\section{The data analysis made by Charlie}
\label{sec:10}

Assume that Alice and Bob meet after the experiment and share their information on all the runs with a new observer Charlie. In particular, all the settings, $a$ or $a'$, respectively $b$ or $b'$ are known, and according to the conditionality principle, all expectations should be calculated conditionally, given these settings. In concrete terms, Charlie has the following data: Settings for Alice in each run, $x_i =a\ \mathrm{or}\ a'$, Settings for Bob, $y_i=b\ \mathrm{or}\ b'$, responses $X_i$ for Alice  ($A$ or $A'$) and $Y_i$ for Bob ($B$ or $B'$), $i=1,...,n$. Charlie wants to do a statistical analysis, and by the conditionality principle he will condition this analysis on all the $x_i$'s and $y_i$'s. His analysis should be concentrated on the following parameters, corresponding to the 4 parts of the data sets that he has received:
\begin{equation}
l_1 =E(XY|x=a,y=b )=E(AB),
\label{Bell6}
\end{equation}
\begin{equation}
l_2 =E(XY|x=a',y=b)=E(A'B),
\label{Bell7}
\end{equation}
\begin{equation}
l_3 =E(XY|x=a,y=b')=E(AB'),
\label{Bell8}
\end{equation}
\begin{equation}
l_4 =E(XY|x=a',y=b')=E(A'B').
\label{Bell9}
\end{equation}

Let us assume a knowledgeable observer Charlie, knowing both statistics and some quantum theory. As any observer, his mind at a given moment can be described by some unit vector in a Hilbert space, which must be taken as big enough to be able to absorb the setting  $a/a'$ and $b/b'$, and the corresponding observations $X$ and $Y$, made by Alice and Bob. This can be accomplished by a four-dimensional Hilbert space, concentrating on the four possible conditional joint distributions of $X$ and $Y$. He should also believe in no-signalling: $E(Z|a,b)=E(Z|a)$ for any variable $Z$ solely connected to Alice.

Put in another way, in order that Charlie should be able to describe any pair $(A=\theta^a, B=\eta^b)$ etc., his Hilbert space $\mathcal{H}'$ must be four-dimensional. We will claim that his state when analysing the data should be seen as an eigenstate of the operator in $\mathcal{H}'$ corresponding to his conceptual variable $\delta =\theta^x \eta^x+\theta^y \eta^y + \theta^z \eta^z$ the dot product of the two inaccessible spin vectors, one belonging to Alice and one to Bob. Even though these spin vectors are inaccessible, $\delta$ is accessible to Charlie.

This can be seen as follows. As analysed in detail in \cite{Ref19}, the operator corresponding to $\delta$ in $\mathcal{H}'$ has two eigenvalues -3 and -1, the single eigenvector corresponding to $\delta=-3$ is given by (\ref{Bell1}), while the eigenspace corresponding to $\delta=-1$ is three-dimensional. Looking at the definition of $\delta$, and the fact that $\theta^x$ and $\eta^x$ etc. all take values $\pm 1$, the value $\delta=-3$ is only possible if $\theta^x \eta^x=\theta^y \eta^y = \theta^z \eta^z =-1$, that is $\eta^x =-\theta^x$ and so on, which implies $\eta^a = -\theta^a$ for all fixed directions $a$. This is just a manification of the fact that Charlie knows that the spin vectors associated with Alice and Bob are equal, but opposite.

Let us assume that before doing any data analysis, Charlie tries to make a probability model for the relevant variables. From the discussion of the previous section, there is a limitation to how much Charlie is able to think of at some given time when making this model. His maximally accessible variables are the different pairs $\zeta=(\theta, \eta)$, where $\theta$ is connected to Alice and $\eta$ is connected to Bob, and both $\theta$ and $\eta$ are binary.

In particular, consider the four pairs $C=(A,B)$, $D=(A,B')$, $E=(A',B)$ and $F=(A',B')$.  Every pair corresponds to one of the 4  parts of the data sets that he is going to analyse. If he thinks hard, he can for instance put up a joint probability model for $C$ and $D$, but this is the maximum of what he is able to do. These two variables are related, so from these two maximally accessible variables he is, according to Theorem 1, able to reconstruct a Hilbert space. Similarly, he is able to construct a Hilbert space from $C$ and $E$; these are related. But the pairs $D$ and $E$ have no relationship to each other; they are essentially different. So we are in the situation of Theorem 2: Charlie is not able to have in his mind all three pairs $C$, $D$ and $E$ when making decisions about the experiment.

We have to verify that the conditions of Theorem 2 hold. Let $\theta=C=(A,B)$ and $\eta=D=(A,B')$. Both are functions of $(\phi ,\psi)$, where $\phi$ is the inaccessible spin vector for Alice's particle, and $\psi$ is the inaccessible spin vector for Bob's particle. As was discussed above, in Charlie's context, each component of $\psi$ is opposite to the corresponding component of $\phi$. This can be written $\psi=-\phi$, and thus both $C$ and $D$ can be seen as functions of $\phi$. It follows then from Lemma 2 that the vector $C=(A,B)$ is permissible with respect to one group, and the vector $D=(A,B')$ is permissible with respect to another group, both seen as functions of $\phi$. The argument needed for finding a group with respect to which both are permissible, goes essentially like the proof of Lemma 1.

From this, the conditions of Theorem 2 hold for $\theta=C$ and $\eta=D$. This implies that Charlie is not simultaneously able to hold in his mind the variable $E$, which is related to $C$, but essentially different from $D$.

 Charlie is thus in particular not able to put up a joint probability model for these 3 pairs. As a consequence, he is not able to put up a joint probability model for the 4 binary variables $AB, AB', A'B'$ and $A'B'$.

Assume now, tentatively, that Charlie is able to put up a joint probability model for his four variables $A, A', B$ and $B'$. Then he would be able to deduce from this model also a joint probability model for the variables $AB, AB', A'B$ and $A'B'$. Thus, from what has just been said, this thesis is impossible. Charlie is not in any way able to think of a joint probability model for his four basic variables. In fact, he is not able to have in his mind all these four variables when making decisions related to the experiment.

This has an important consequence: From this point of view, Charlie is simply not able to follow the arguments leading to (\ref{Bell2}) and then to (\ref{Bell3}) when making his decisions. Thus he can not then follow the arguments leading to the CHSH inequality, and must see the validity of this inequality either as an empirical question or a question that can be resolved by his knowledge about quantum mechanics.

From his data, Charlie can compute natural estimates: $\widehat{l_1}=\overline{AB}$, $\widehat{l_2}=\overline{A'B}$, $\widehat{l_3}=\overline{AB'}$ and $\widehat{l_4}=\overline{A'B'}$. Let us further assume that the settings are such that, by the Born formula, which gives $E(AB)=-\mathrm{cos}(a,b)$, the CHSH inequality is violated. (Again, one choice, in some sense the optimal one, is $a\sim 0^o$, $a'\sim 90^o$, $b\sim 225^o$ and $b'\sim 135^o$.) Then, if the number $n$ of runs is large enough, Charlie will find by using Born's formula before looking at his data, that with high probability from this:
\begin{equation}
\overline{AB}+\overline{A'B}+\overline{AB'}-\overline{A'B'} > 2.
\label{Bell10}
\end{equation}

This may convince him that the CHSH inequality is \emph{not} valid, and he will be surprised if his estimates do not satisfy (\ref{Bell10}). He will also predict that for a future series of runs, if the number of runs is large enough and the settings are as before, then (\ref{Bell10}) will hold with large probability.

Parts of this discussion does not depend on Charlie's possible knowledge about quantum theory, but it is all the time assumed that he is in a decision situation which, as described in \cite{Ref21} can be connected to a finite number of prospects. 

In his modelling effort Charlie might be able to find separate numbers in separate models for $l_1 =E(AB)$, $l_2=E(A'B)$, $l_3=E(AB')$ and $l_4=E(A'B')$. These numbers may be compared either to the prediction made by quantum theory, or to the estimates he found from the data from Alice and Bob. In either case, given suitable settings $a, a', b$ and $b'$, he may be convinced that the CHSH inequality may be violated.

Thus, by Bell's theorem he believes that the assumption of local realism must be violated. He may be convinced about the validity of Einstein's relativity theory, and from this he may deduce that the locality assumption  should hold. Hence his only option is to reject the universal assumption of realism. Charlie may be convinced of the statement advocated in \cite{Ref25}: The only `real things' in physics are events, and any theory, any set of questions to be answered, should be connected to the perception of events made by an actor or by a group of communicating actors. As argued here, Charlie as an actor is limited.

\section{Charlie, Alice and others}
\label{sec:11}

Assume that Alice and Bob plan to make a new Bell experiment with $n$ runs  together.
Assume also that Alice and Charlie meet and talk together after Charlie has done a data analysis, but before any new series of runs. The issue of their talk is the predicton (\ref{Bell10}) for the new experiment. Charlie may be convinced about this prediction, but Alice is still unsure.

Another situation might be that Charlie chooses to market his prediction to all people that he knows, telling them about his arguments behind this prediction. This must be seen as a form of information processing. In a recent paper \cite{Ref24} a model for joint decisions made by a group of people is based on an extensive exchange of information. Such an analysis may then consider possible future joint decisions made either by Charlie and Alice together or by Charlie and his friends together.

I will not go into details here, but concentrate on the following: The discussion between Alice and Charlie focuses on the single binary variable $\zeta = \mathrm{sign} (\overline{AB}+\overline{A'B}+\overline{AB'}-\overline{A'B'} - 2)$ for the new experiment. According to \cite{RefJ8}, Alice's question to Charlie on this variable may be seen as a measurement. If she should accept Charlie's answer, she will enter a new state partly given by $\zeta = +1$ for the relevant set of settings $a, a', b$ and $b'$ in the future Bell experiment. She will believe that an empirical version of the CHSH inequality will be violated in the new experiment if $n$ is large enough.  And she may be convinced by Charlie's arguments around local realism.

The discussion with other people may be more complicated. But if Charlie's arguments are strong enough, both theoretical arguments from quantum mechanics and empirical results, most of his friends will probably enter a state partly given by $\zeta =+1$: Thus they will with some part of their mind believe that the CHSH inequality may be violated under suitable conditions. Hence in the light of Bell's theorem they will not be convinced that local realism, made precise in some way, will always hold.

The setting for this last discussion must be such that the friends can communicate with Charlie. This may be argued to imply that all the friends have some relation to a four-dimensional Hilbertspace in connection to one run of the Bell experiment, and that their mental state here then is given by (\ref{Bell1}), that is, corresponding to $\delta=-3$. Then, by the above arguments, neither of them will then have the possibility to, at the same time, have all the variables $A, A', B$ and $B'$ in their mind, and thus they will not be convinced of the argument leading to (\ref{Bell2}).

\section{Discussion}
\label{sec:12}

We are all limited. Like Alice and Charlie neither of us can always answer all questions, even simple ones that require a yes/no answer. This is of course obvious, but one aspect of this may not be clear to everybody: Our mind is limited by how many conceptual variables we can think of at the same time when making a decision. Like Alice, we can sometimes seek answers from people having more insight.

Decisions may be made by single a actor or by groups of communicating actors. In some situations these decisions may be related to measurements that we are about to make, and these measurements can be formulated by focused questions to nature involving accessible conceptual variables. This is the point of departure for the approach to quantum mechanics given in \cite{RefB1}.

Going back to Sect. \ref{sec:10}, look at Charlie's efforts to make a probability model over his variables. In the language of statisticians \cite{RefB1}, these probabilities may be called epistemic probabilities. He is able to make joint probability models over pairs of variables, but not over all 4 variables. This must mean that these epistemic models in general are different than ordinary probability models. The fact that quantum probabilities behave differently than ordinary probabilities is well known \cite{Ref26}.

\section{Conclusions}
\label{sec:13}

The discussions around Bell's theorem and the assumptions of local realism may perhaps continue. In my view the paradoxes around this issue may be resolved by considering an actor like Charlie discussed above. After analysing his Bell experiment data, he is convinced that the assumption of local realism cannot hold in general. He has two arguments for this: Empirical results and a belief on the general validity of quantum mechanics. At the same time he is not able to follow the simple arguments leading to the CHSH inequality. As I see it, this may be said to be so because of his limitation: He is simply not able to keep enough variables at the same time in his mind when making his decisions.

My arguments in this paper has partly rested on the epistemic process approach towards quantum theory \cite{RefB1}. However, the arguments concentrated on the actor Charlie also seem to have some universal validity. It must be concluded that local realism is not longer a universally convincing position, given these arguments. And this can be highlighted by focusing on the world as seen by any specific actor or by a group of different, communicating, actors, all being limited in the specific sense discussed in Sect. \ref{sec:11} and Sect. \ref{sec:12} above. 

The process of making decisions may in certain situations be a difficult one. In this paper I have argued that individual decisions are dependent on conceptual variables in the mind of the person who makes the decisions. Other decisions are made more rutinely, being determined by our upbringing and by the social context that we live in. In some of our decisions, we are inspired by other people. In our Western culture, religious faith or related, deep issues, may alo play an important part. My own thoughts around this is given in \cite{Ref81}.

In practice many decisions are made jointly by groups of people that communicate. Also in the latter case a common philosophy and through this, common conceptual variables play a role. Decisions may be initiated by persons that the group look up to. In this world, really cruel and disastrous decisions have also been made in this way. In a civilizest society such decisions should be countered. In an idealized world, all decisions should be made in a way that is rational and at the same time has a high ethical standard. Also, in this ideal world, science in all its variants should lead the way here.

\begin{acknowledgements}
I am grateful to Chris Ennis and to Peter Morgan for discussions.
\end{acknowledgements}

%
%



\bigskip

\textbf{Appendix. Operators and their properties.}

In \cite{Ref50} I took as a point of departure Chapter 2 in Perelomov \cite{Ref60}, which discusses coherent states for arbitrary Lie groups. Let $G$ be a transitive group acting on the space $\Omega_\theta$ associated with some conceptual variable $\theta$ and $U(g)$ its unitary irreducible representation acting on the Hilbert space $\mathcal{H}$. I will assume that $G$ has a trivial isotropy group, so that the elements $g$ of $G$ are in one-to-one correspondence with the values of $\theta$.

As in \cite{Ref50} (and in \cite{RefB1}) take a fixed vector $|\theta_0\rangle$ in $\mathcal{H}$, and consider the set $\{|\theta\rangle\}$, where $|\theta\rangle = U(g)|\theta_0\rangle$ with $g$ corresponding to the value $\theta$. It is not difficult to see that two vectors $|\theta^1\rangle$ and $|\theta^2\rangle$ correspond to the same state, i.e., differ by a phase factor ($|\theta^1\rangle =\mathrm{exp}(i\alpha)|\theta^2\rangle$, $|\mathrm{exp}(i\alpha)| =1$), only if $U(g^{2, -1}g^1)|\theta_0\rangle=\mathrm{exp}(i\alpha)|\theta_0\rangle$, where $g^1$ corresponds to $\theta^1$ and $g^2$ corresponds to $\theta^2$. Suppose $E=\{e\}$ is a subgroup of the group $G$, such that its elements have the property

\begin{equation}
U(e)|\theta_0\rangle =\mathrm{exp}[i\alpha(e)]|\theta_0\rangle .
\label{u1}
\end{equation}

When the subgroup $E$ is maximal, it will be called the isotropy subgroup for the state $|\theta_0\rangle$. More precisely, it is the isotropy subgroup of the group $U(G)$ corresponding to this state.

The construction shows that the vectors $|\theta\rangle$ corresponding to a value $\theta$ and thus to an element $g\in G$, for all the group elements $g$ belonging to a left coset class of $G$ with respect to the subgroup $E$, differ only in a phase factor and so determine the same state. Choosing a representative $g(x)$ in any equivalence class $x$, one gets a set of states $\{|\theta_{g(x)}\rangle\}$, where $x\in X=G/E$. Again, using the correspondence between $g$ and $\theta$, I will write these states as $\{|\theta(x)\rangle\}$, or in a more concise form $\{|x\rangle\}$, $|x\rangle\in \mathcal{H}$. 
\bigskip

\textbf{Definition A1} \textit{The system of states $\{|\theta\rangle =U(g)|\theta_0\rangle\}$, where $g$ corresponds to $\theta$ as above, is called the coherent-state system $\{U,|\theta_0\rangle\}$. Let $E$ be the isotropy subgroup for the state $|\theta_0\rangle$. Then the coherent state $|\theta(g)\rangle$ is determined by a point $x=x(g)$ in the coset space $G/E$ corresponding to $g$ and to $|\theta(g)\rangle$ is defined by $|\theta(g)\rangle=\mathrm{exp}(i\alpha)|x\rangle$, $|\theta_0\rangle=|0\rangle$.}
\bigskip

\textbf{Remark.} The states corresponding to the vector $|x\rangle$ may also be considered as a one-dimensional subspace in $\mathcal{H}$, or as a projector $\Pi_x =|x\rangle\langle x|$, $\mathrm{dim} \Pi_x=1$, in $\mathcal{H}$. Thus the system of coherent states, as defined above, determines a set of one-dimensional subspaces in $\mathcal{H}$, parametrized by points of the homogeneous space $X=G/E$.
\bigskip

In \cite{Ref50} this theory is generalized to the case with two maximally accessible conceptual variables $\theta$ and $\eta$ connected through an underlying inaccessible variable $\phi$ by a transformation $k$ of the underlying space $\Omega_\phi$ by $\eta(\phi)=\theta(k\phi)$. For this case define $\psi=(\theta,\eta)$. It is shown in \cite{Ref50} that a group $N$ can be defined on $\psi$ given by the group $G$ above and its isomorphic `copy' $H$ acting on $\eta$. Let $M$ be the isotropy subgroup of $N$ corresponding to $E$ above, and define the coset $Z=N/M$. Then it is shown in \cite{Ref50} that there is an irreducible representation $W(\cdot)$ acting on $N$, formed by the representation $U(\cdot)$ of $G$ and the corresponding representation $V(\cdot)$ of $H$. These are acting on the same Hilbert space $\mathcal{H}$, and this implied states $|\psi(\phi)\rangle=W(n)|\psi_0\rangle$ constructed as above, defined on the same Hilbert space, and that they satisfy a resolution of the identity

\begin{equation}
\int |\psi\rangle\langle\psi |\nu (d\psi)=I
\label{u00}
\end{equation}
for a left-invariant measure $\nu$ on $\Psi = \{\psi\}$. 

Furthermore, it is shown that $z=(x,y)$, where $x$ is an element of $X$ and $y$ is an element of the corresponding coset $Y=H/F$. Here $F$ is the subgroup of $H$ corresponding to the subgroup $E$ of $G$.

In fact, the situation here is completely symmetric between $(\theta, G, E, X)$ on the one side and $(\eta, H, F, Y)$ on the other side. Furthermore,  the transformations $N$ on $\psi=(\theta,\eta)$ are constructed from independent transformations in $G$ on $\theta$ and transformations in $H$ on $\theta$. Then it is reasonable to assume that the measure $\nu$ in (\ref{u00}) can be written as $\nu(d\psi )=\rho(dx)\rho(dy)$ for some marginal measure $\rho$.

From this, the operators corresponding to $\theta$ and $\eta$ can be defined by

\begin{equation}
P(x)=\int_Y  |\psi\rangle\langle\psi| \rho(dy),
\label{u6}
\end{equation}

and

\begin{equation}
A^\theta = \int_X \theta(x) P(x) \rho (dx).
\label{u7}
\end{equation}

Similarly:

\begin{equation}
Q(y)=\int_X  |\psi\rangle\langle\psi| \rho(dx),
\label{u8}
\end{equation}

and

\begin{equation}
A^\eta = \int_Y \xi(y) Q(y) \rho (dy).
\label{u9}
\end{equation}

Here, to recall, $y$ is defined as an element of the homogeneous space $Y=H/F$, where $H$ is a transitive group, isomorphic to $G$, acting on the space $\Omega_\eta$ on which $\eta$ varies, and $F$ is the subgroup of $H$ corresponding to $E$ of $G$.

For conceptual variables $\xi$ that are not maximally accessible, we can always write $\xi=f(\theta)$ for some maximal $\theta$, and an operator for $\xi$ can be found by using the spectral theorem on the operator $A^\theta$ (cp. equation (4.30) in \cite{RefB1}). 
\bigskip

To illustrate the theory, I will prove an important result, a correction/ precision of Theorem 4.2 in \cite{RefB1}. Then first I have to define the notion of permissibility.
\bigskip

\textbf{Definition A2}  \textit{The function $\theta(\cdot)$ on a space $\Omega_\phi$ upon which a group of transformations $K$ is defined, is said to be permissible if the following holds: $\theta(\phi_1)=\theta(\phi_2)$ implies $\theta(k\phi_1)=\theta(k\phi_2)$ for all $k\in K$.}
\bigskip

This notion is studied thoroughly in \cite{Ref61}. The main conclusion is that if $\theta(\cdot)$ is permissible, then there is a group $G$ acting on the image space $\Omega_\theta$ such that $g(\theta(\phi))$ is defined as $\theta(k\phi)$; $k\in K$. The mapping from $K$ to $G$ is an homomorphism. If $K$ is transitive on $\Omega_\phi$, then $G$ is transitive on $\Omega_\theta$. (Lemma 4.3 in \cite{RefB1}.) It is easy to show that $G$ has a trivial isotropy group if $K$ has a trivial isotropy group.
\bigskip

\textbf{Theorem A1}
\textit{Assume that the function $\theta(\cdot)$ is permissible with respect to a group $K$ acting on $\Omega_\phi$. Assume that $K$ is transitive and has a trivial isotropy group. Let $T(\cdot)$ be a unitary representation of $K$ such that the coherent states $T(k)|\psi_0\rangle$ are in one-to-one correspondence with $k$. For any transformation $t\in K$ and any such unitary representation $T$ of $K$, the operator $T(t)^\dagger A^\theta T(t)$ is the operator corresponding to $\theta'$ defined by $\theta'(\phi)=\theta(t\phi)$.}

\bigskip

\textit{Proof.}
By (\ref{u7}) we have
\begin{equation}
T(t^{-1})A^\theta T(t)=\int_X \theta(x) P_t(x)\rho (dx),
\label{u10}
\end{equation}
where
\begin{equation}
P_t(x)=\int_{Y} |\psi(t^{-1}\phi)\rangle\langle \psi(t^{-1}\phi)| \rho (dy).
\label{u11}
\end{equation} 

To show this, we need to prove that $T(t^{-1}) |\psi(\phi)\rangle =|\psi (t^{-1}\phi )\rangle$.

Note that $t^{-1}\phi$, permissibility and $\theta=\theta(\phi)$, $\eta=\eta(\phi)$ induces from $t$ a new transformation $s$ acting on $\psi=(\theta ,\eta)$. Consider $s^{-1}(\theta ,\eta )=s^{-1}(g,h)$ from the one-to-one correspondence between $\theta$ and $g$ and between $\eta$ and $h$. Let $S(s^{-1})=T(t^{-1})$ under this correspondence. Then $T(t^{-1}) |\psi(\phi)\rangle =S(s^{-1})|\psi\rangle$, where $|\psi\rangle=T(k)|\psi_0\rangle$ is the vector i $\mathcal{H}$ which is in one-to-one correspondence with the group element $k$, which generates $g=g(k)$ and hence $\theta$ by permissibility. Now the group $H$, which corresponds to a copy of $G$, can also be seen as generated by $K$, so $h=h(k)$. This implies that $(g,h)$ is a function of $k$, and by permissibility also $\psi=(\theta, \eta)$ is a function of $k$. Since the two maximally accessible variables $\theta$ and $\eta$ determine the Hilbert space, this function must be one-to-one.
 
 Write the group elements $s^{-1}(g,h)$ as $(g',h')$, new members of groups $G$ and $H$ acting on $\theta$ and $\eta$, respectively. Let again $E$ be the subgroup of $G$ constructed as in (\ref{u1}), and let $F$ be the corrsponding subgroup of $H$. Write $X=G/E$ and $Y=H/F$. The new elements of these cosets may be defined as $x'=t^{-1}x$ and $y'=t^{-1}y$, respectively. This gives an element $z'=(x',y')$, and the corresponding state in $\mathcal{H}$ as $|z'\rangle$, which as in Definition A1 also, modulus a phase factor, can be written as $|\psi(g',h')\rangle=|\psi(t^{-1}\phi)\rangle$.
 
 Hence (\ref{u10}) and (\ref{u11}) can be written as

\begin{equation}
T(t^{-1})A^\theta T(t)=\int_X \theta(x) P_t(x)\rho (dt^{-1}x),
\label{u12}
\end{equation}
where
\begin{equation}
P_t(x)=\int_{Y} |\psi(t^{-1}\phi)\rangle\langle \psi(t^{-1}\phi)| \rho (dt^{-1}y).
\label{u13}
\end{equation} 

Now make a change of variables from $(x,y)$ to $(x',y')=(tx,ty)$ in these integrals. Since $\nu(d\psi)=\rho (dx)\rho(dy)$ is left invariant, the corresponding $\rho$ may be taken to be left invariant. Therefore the last integral may be written

\begin{equation}
P_t(x)=\int_{Y} |\psi(\theta(t^{-1}\phi), \eta(\phi')\rangle\langle \psi(\theta(t^{-1}\phi), \eta(\phi')| \rho (dy'),
\label{u14}
\end{equation} 
and we can write $P_t(x)=P(t^{-1}x)$. This is inserted into (\ref{u12}), and using left-invariance of the measure again, this gives that the operator $T(t^{-1})A^\theta T(t)$ is associated with the conceptual variable $\theta (tx)$, which also may be written as $\theta(t\phi)$.
\qed
\bigskip

By using this result in the same way as Theorem 4.2 is used in \cite{RefB1}, a rich theory follows. I will limit me here to the case where $\theta$ is a discrete conceptual variable. Then one can show:

1) The eigenvalues of $A^\theta$ coincide with the values of $\theta$.

2) The variable $\theta$ is maximally accessible if and only if the eigenvalues of $A^\theta$ are non-degenerate.

3) For the maximal case the following holds in a given context: a) For a fixed $\theta$ each question `What is the value of $\theta$?' together with a sharp answer `$\theta=u$' can be associated with a normalized eigenvector of the corresponding $A^\theta$. b) If there in the context are a set $\{\theta^a ; a\in \mathcal{A}\}$ of maximally accessible conceptual variables (these must by the results of Section 8 be related to each other) one can consider all ket vectors that are normalized eigenvectors of some operator $A^{\theta^a}$. Then each  of these may be associated with a unique question-and-answer as above.

\end{document}